\documentclass[12pt]{article}
\usepackage{graphics,cite,amssymb,epsfig,float}
\usepackage[usenames,dvips]{color}

\newcommand{\lsim}{\raisebox{-0.13cm}{~\shortstack{$<$ \\[-0.07cm] $\sim$}}~} 
\newcommand{\gsim}{\raisebox{-0.13cm}{~\shortstack{$>$ \\[-0.07cm] $\sim$}}~}

\newcommand{\beq}{\begin{eqnarray}} 
\newcommand{\eeq}{\end{eqnarray}} 
\newcommand{\s}{\smallskip} 

\begin{document}
\baselineskip=16pt 

\rightline{LPT-ORSAY-08-26}
\rightline{\today}    
 
\vspace{0.7cm} 
 
\begin{center} 
 
{\Large{ \bf The constrained next-to-minimal supersymmetric}}

\vspace{0.2cm} 

{\Large{ \bf  standard model}}

\vspace{0.7cm} 

{\sc A. Djouadi$^{1,2,3}$, U. Ellwanger$^1$} and {\sc A.M. Teixeira$^1$} 

\vspace{0.7cm} 

$^1$ Laboratoire de Physique Th\'eorique, CNRS and Universit\'e Paris--Sud,\\ 
F--91405 Orsay Cedex, France.\\ 
$^2$ School of Physics and Astronomy, Southampton University, 
Highfield, SO17 1BJ, UK.\\ 
$^3$ Physikalisches Institute, University of Bonn, Nussallee 12, 
53115 Bonn, Germany.

\end{center} 

\vspace{0.5cm}  

\begin{abstract} 

We consider the fully constrained version of the next-to-minimal supersymmetric
extension of the standard model (cNMSSM) in which a singlet Higgs superfield is
added to the two doublets that are present in the  minimal extension (MSSM).
Assuming universal boundary conditions at a high scale for the soft
supersymmetry-breaking gaugino, sfermion and Higgs mass parameters as well as
for the trilinear interactions, we find that the model is more constrained than
the celebrated minimal supergravity model. The phenomenologically viable region
in the parameter space of the cNMSSM corresponds to a small value for the
universal scalar mass $m_0$: in this case, one single  input parameter is
sufficient to describe the phenomenology of the model once the available
constraints from collider data and cosmology are imposed. We present the
particle spectrum of this very predictive model and discuss how it can be 
distinguished from the MSSM.

\end{abstract} 

\newpage

The next-to-minimal supersymmetric standard model (NMSSM) \cite{genNMSSM1,
genNMSSM2,genNMSSM3}, in which the spectrum of the minimal extension (MSSM) is
extended by one singlet superfield, was among the first  supersymmetric (SUSY)
models based on supergravity-induced SUSY-breaking terms. It has gained a
renewed interest in the last decade, since it solves in a natural and elegant
way the so-called $\mu$ problem~\cite{MuProblem} of the MSSM; in the NMSSM this
parameter is linked to the vacuum expectation value (vev) of the singlet Higgs
field, generating a value close to the SUSY-breaking scale. Furthermore, it
leads to an interesting phenomenology as the MSSM spectrum is extended to
include an additional  CP-even and CP-odd  Higgs state as well as a fifth
neutralino, the singlino. \s

In contrast to the non- or partially constrained versions of the NMSSM that have
been intensively studied in the recent years \cite{benchmark}, and which involve
many free parameters, the constrained model (cNMSSM) has soft SUSY-breaking
para\-meters that are universal at a high  scale. This is motivated  by schemes
for SUSY-breaking that are mediated by (flavor blind)  gravitational
interactions, and leads to a more predictive model as the number of unknown
parameters is reduced to a handful. \s

In the constrained MSSM or minimal supergravity (mSUGRA)  scenario
\cite{msugra},  the universality of the soft parameters at the grand unification
scale $M_{\rm GUT}$ leads to only four continuous parameters: the common
gaugino  $M_{1/2}$ and   scalar $m_0$ masses, the trilinear coupling $A_0$, and
the ratio of vevs of the two Higgs fields $\tan\beta$. (The conditions for a
correct electroweak symmetry breaking allow to trade the two basic Lagrangian
parameters $\mu^2$ and  its  corresponding soft breaking term $B\mu$ by the $Z$
boson mass $M_Z$ and  $\tan\beta$, leaving the sign of $\mu$ undetermined.) The
numerous soft parameters at low energies are then obtained through
renormalization group (RG)  evolution. The \nobreak{cNMSSM} with universal soft
terms at the GUT scale has the same number of unknown parameters: instead of the
cMSSM $\mu$ and $B$ parameters, one has the trilinear couplings $\lambda$ and
$\kappa$ in the  Higgs sector. \s

General features of the cNMSSM parameter space as well as its  phenomenology
have been discussed in Refs.~\cite{genNMSSM2,genNMSSM3}. These studies already
revealed that the allowed range for the parameters $M_{1/2}, m_0$ and $A_0$ is
different from that of the cMSSM. Indeed while small values for $m_0$ are
disfavored  in the cMSSM as they lead to charged sleptons that are lighter than
the neutralino $\chi_1^0$, which is the preferred lightest SUSY particle (LSP),
small $m_0$ is needed in the cNMSSM to generate a non-vanishing vev of the
singlet  Higgs field \cite{genNMSSM2}; the slepton LSP problem can be evaded
owing to the presence of the additional singlino-like neutralino which, in
large  regions of the cNMSSM parameter space, is the true LSP. Note that
vanishing values for $m_0$ are  naturally obtained in supergravity models with
K\"ahler potentials of the no-scale type \cite{noscale}; however, the additional
no-scale  prediction $A_0 =0$  is difficult to realize in the cNMSSM.\s

Since the early studies, bounds on the Higgs and SUSY particle spectrum from
high-energy collider data and low-energy measurements have become more severe,
while important inputs such as the top quark mass are more accurately measured
\cite{pdg}. In addition, tools for a more precise determination of the mass
spectrum and couplings \cite{nmssmtools}, and the cosmological dark matter
relic  density \cite{belsemenov} have become available. In this letter, we
re-investigate the parameter space of the cNMSSM in the light of these recent
constraints, using the updated tools.\s

We find a viable region of the cNMSSM parameter space very close to the
no-scale  situation: vanishing or very small $m_0$ and $|A_0|$ well below
$M_{1/2}$. In fact, in order to obtain the correct cosmological relic density
for the LSP, which turns out to be a singlino-like neutralino \cite{hugbelpuk},
$A_0$ is essentially fixed in terms of $M_{1/2}$; $\tan\beta$ is determined by 
the scalar mass universality (including the additional singlet), while LEP
constraints on the Higgs sector imply $\lambda \lsim 10^{-2}$ and this parameter
has practically no effect on the remaining particle spectrum. \s

Thus, remarkably, only {\it one single parameter} is sufficient to describe the
phenomenological features of the cNMSSM. This makes the model much more
predictive than the celebrated mSUGRA model. In addition, as will be shown, the
phenomenology differs considerably in the two scenarios.\s

In this analysis, we consider the NMSSM with a scale invariant superpotential 
given by 
\begin{equation}  
{\cal W}\!=\!\lambda {S} {H}_u {H}_d + \frac{\kappa}{3}\, S^3 + \dots   
\label{supot}  
\end{equation} 
where the two terms shown substitute the $\mu H_u H_d$ term in the MSSM
superpotential and we have omitted the usual generalization of the Yukawa
interactions. The soft SUSY-breaking terms consist of mass parameters for the
gauginos $M_{1,2,3}$, sfermions $m_{\tilde F_{L,R}}$ and Higgs fields
$m_{H_{u,d}}$ and trilinear interactions $A_f$ as in the MSSM, supplemented by 
an additional scalar mass $m_S$ and two trilinear couplings $A_\kappa$ and 
$A_\lambda$ for the singlet field. Once the unification of these soft 
SUSY-breaking masses and trilinear couplings at the scale $M_{\rm GUT}$ is 
imposed, the Lagrangian of the cNMSSM depends on the five parameters,
$M_{1/2},m_0,  A_0, \lambda$ and $\kappa$. The  correct value for $M_Z$ reduces
the dimension of the  parameter space from five to four; e.g. $\kappa$ can be
determined in terms of the other parameters. Hence, the  number of continuous
free parameters in the cNMSSM  is the same as in the cMSSM.\s

For practical purposes it is convenient to adopt the following procedure: 
In addition to $M_Z$, one allows for the five cNMSSM input parameters
\beq
M_{1/2}\ , \ m_0 \ , \ A_0 \ , \ \lambda \ \ {\rm and} \  \tan\beta\;.
\eeq
The parameters $\kappa$ and the soft singlet mass squared $m_S^2$ are both  
determined at the scale $M_{\rm SUSY}$ through the minimization equations of
the  scalar potential. (The vev $\left<S \right>$ or $\mu_{\rm eff}\equiv
\lambda  \left<S \right>$ is also fixed through the third independent
minimization equation, which leaves the sign of $\mu_{\rm eff}$ undetermined.)
This is the procedure employed by the routine NMSPEC within NMSSMTools
\cite{nmssmtools}, which calculates the spectra of the Higgs and  SUSY
particles  in the NMSSM in terms of the soft SUSY-breaking terms at $M_{\rm
GUT}$ (except for the parameter $m_S^2$), $\tan\beta$ at the weak scale and
$\lambda$ at the SUSY-breaking  scale $M_{\rm SUSY}$. \s

Clearly, the soft singlet mass squared $m_S^2$ at $M_{\rm GUT}$ will in
general   not coincide with $m_0^2$. However, one can confine oneself to regions
in parameter space where the difference between $m_S^2$ and $m_0^2$ is
negligibly small. This condition leaves us with an effective four-dimensional 
parameter space, consistent with the considerations above.\s

Let us now explore the cNMSSM parameter space defined by arbitrary values of
$M_{1/2}$, $A_0$ and $\lambda$, assuming $\mu_{\rm eff}$ positive and,  for the
time being, restricting  ourselves  to $m_0 = 0$.  For each choice of $M_{1/2}$,
$A_0$ and $\lambda$, $\tan\beta$ is determined by the requirement that $m_S^2$
at $M_{\rm GUT}$ should be close to $m_0^2 = 0$. In  practice, we impose
$m_{S}^2(M_{\rm GUT}) < (5~{\rm GeV})^2$, which typically requires  to tune the
fourth decimal of $\tan\beta$ (this should not be interpreted as a finetuning,
since $m_S^2$ should be considered as an input parameter, whereas $\tan\beta$ is
determined by the minimization of the effective potential).  For the most
relevant SM parameters we chose $\alpha_s(M_{Z}) = 0.1172$, $m_b(m_b)^{ 
\overline{\rm MS}} = 4.214$ GeV and $m_{\mathrm{top}}^{\mathrm{pole}} = 171.4$
GeV \cite{pdg}.\s

For this set of parameters, we select the cNMSSM space which survives once one
imposes theoretical requirements such  as  correct electroweak symmetry
breaking,  perturbative couplings at the high scale, the absence of tachyonic
masses, a neutralino LSP, etc.., and phenomenological constraints such as the
LEP bounds on Higgs masses and couplings, collider bounds on the SUSY particle
masses \cite{pdg}, experimental data from $B$-meson physics \cite{bphys} and
from the anomalous magnetic  moment of the muon \cite{g-2}, and a relic density
compatible with cosmological data \cite{wmap}. For $\lambda \lsim  10^{-2}$, as
it turns out to be the case (see below), the phenomenologically allowed region
is nearly independent of the input $\lambda$. \s

Leaving aside, for the time being, the WMAP constraints on the LSP relic
density  $\Omega h^2$,  but including all the other  constraints above, the
phenomenologically allowed region in the $[M_{1/2},A_0]$  plane is shown in
Fig.~1.\s

\begin{figure}[ht]
\begin{center}
\includegraphics[width=0.5\linewidth,angle=-90]{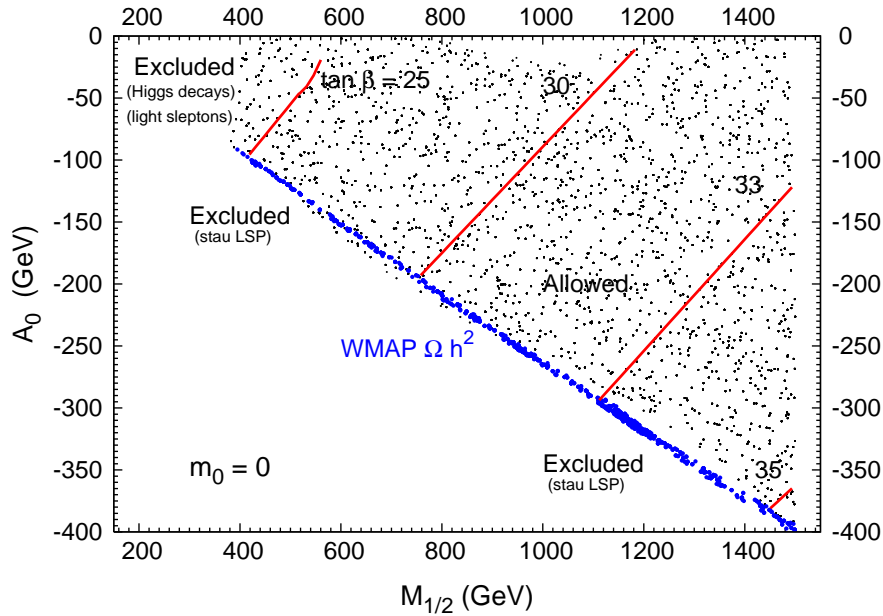}
\end{center}
\vspace*{-2mm}
\caption{The viable cNMSSM region in the $[M_{1/2},A_0]$  plane for $m_0 \sim 
0$ and $\lambda=2 \times 10^{-3}$, once  theoretical, collider and cosmological
constraints have been imposed.}
\end{figure}

This allowed region is bounded from below, i.e. for large absolute values of
$|A_0|$, by the absence of a charged (generally tau slepton) LSP as in the
cMSSM  for $m_0=0$; inside the allowed region, the LSP is a singlino-like state.
The upper bound at $A_0 \sim 0$ follows from the positivity of the mass squared
of the singlet-like CP-odd Higgs boson, which is given to a good approximation
by $-3\kappa A_\kappa \left< S\right>$, $\kappa\left< S\right>$ being positive, 
and $A_\kappa \sim A_0$. To the left, i.e. towards smaller values of $M_{1/2}$,
the allowed region is bounded simultaneously by the condition that the lightest
tau slepton mass must be above $\sim$~100~GeV from its non-observation at LEP,
and the mass of the lightest SM-like CP-even Higgs boson above $\sim$ 114 GeV.
To the right, i.e. towards larger values of $M_{1/2}$ for which SUSY particles
are too heavy to be produced at the LHC, no ``physical'' constraint appears. We
believe, however, that this region requires more and more hidden finetuning
\footnote{The actual finetuning is hidden by our procedure which determines
$\kappa$ in terms of other parameters such as $M_Z$; this is similar to the
determination of $B$ in the cMSSM.}.  Inside this allowed region, $\tan\beta$ is
fixed by the requirement $m_S^2 (M_{\rm GUT}) \sim 0$, and  turns out to be
quite large (see Ref.~\cite{largetb} for earlier work on the cNMSSM at large
$\tan\beta$).  In Fig.~1,  we have indicated lines corresponding to constant
$\tan\beta = 25, 30, 33$ and 35.\s 

As a next step, we require that the  WMAP constraint on the relic density of
the  $\chi_1^0$ dark matter (DM) candidate, calculated using the program 
MicrOMEGAS \cite{belsemenov}, is satisfied. Given the actual small error bars,
$0.094 \, \lesssim \Omega_{\chi_1^0} h^2 \lesssim 0.136$ at the $2\sigma$ level
\cite{wmap}, this constraint (if satisfied at all)  reduces the parameter space
of  any model to a lower dimensional hypersurface. Within the allowed region in
Fig.~1, the  correct relic density for $\chi_1^0$ is obtained along the  line
close to the lower boundary (DM line), where the mass of the singlino-like LSP
is close to  the mass of the next-to-LSP which is the lightest tau slepton
$\tilde \tau_1$, $M_{\tilde \tau_1} - M_{\chi^0_1} \sim$ 3 to 5~GeV. Near the
upper  boundary, the mass of the singlino-like LSP becomes very small, implying
a far too large relic density. For $\lambda \ll 1$ (see below), the
singlino-like $\chi_1^0$ LSP  has very small couplings to SM particles and,
thus, the main process which generates the correct relic density is the assisted
annihilation or ``co-annihilation" of tau sleptons, $\tilde \tau_1  \tilde
\tau_1 \to$ SM particles, which does not depend on these
couplings\footnote{However,   $\lambda$ values larger than $\sim 10^{-5}$,  as
used here, are needed such that the hypothesis of thermal equilibrium between
the LSP and the next-to-LSP near  the relevant temperature can be considered
satisfactory.}.\s

As mentioned previously, we have checked that constraints from $B$-physics 
\cite{bphys}, such as the branching ratio for the radiative decays $b \to
s\gamma$,  are satisfied. Moreover, we find that the supersymmetric
contribution  to the anomalous magnetic moment of the muon,  $\delta a_\mu=
(g-2)_\mu$,   accounts for the $\sim 3\sigma$ deviation from the SM 
expectation~\cite{g-2}:  along the DM  line, $\delta a_\mu$ decreases from $\sim
4.2\times 10^{-9}$ for $M_{1/2} \sim 400$~GeV to  $\sim 0.2\times 10^{-9}$ for
$M_{1/2} \sim 1.5$~TeV. In view of a desired value $\delta a_\mu\sim (2.7\pm
2)\times 10^{-9}$, the region $M_{1/2} \lsim 1$~TeV is thus preferred by this
observable.\s 

At this  stage, one can conclude that for small $m_0$ and $\lambda$, only one
single parameter is needed to describe the phenomenology of the cNMSSM, once 
collider and cosmological constraints are imposed. Along the  DM line in
Fig.~1,  once a value for $M_{1/2}$ is chosen, both the universal trilinear
coupling $A_0$ and $\tan\beta$ are uniquely fixed.  Before investigating the
impact of other values for the parameters $m_0$ and $\lambda$, let us first
discuss the Higgs and sparticle spectrum as a function of $M_{1/2}$ in the
phenomenologically allowed region.  The Higgs, neutralino and stau masses are
shown in  Fig.~2 where we also indicate the corresponding values of $A_0$.\s 

The essential features of the Higgs spectrum are as follows. For $M_{1/2} \lsim
660$~GeV, the lightest CP-even Higgs boson has a dominant singlet component,
hence a very small coupling to the $Z$ boson, which allows it to escape LEP
constraints. The next-to-lightest CP-even scalar is SM-like, with a mass
slightly above 114~GeV. The lightest CP-odd scalar is again singlet-like, with
a  mass above $\sim 120$~GeV. The heaviest CP-even and CP-odd scalars are
practically degenerate with the charged Higgs boson, with masses above $\sim
520$ GeV. For $M_{1/2} \gsim 660$~GeV, the lightest CP-even scalar is SM-like
with a mass increasing slightly with $M_{1/2}$ up to $\sim 120$~GeV, while the
next-to-lightest one is now singlet-like.\s

\begin{figure}[!ht]
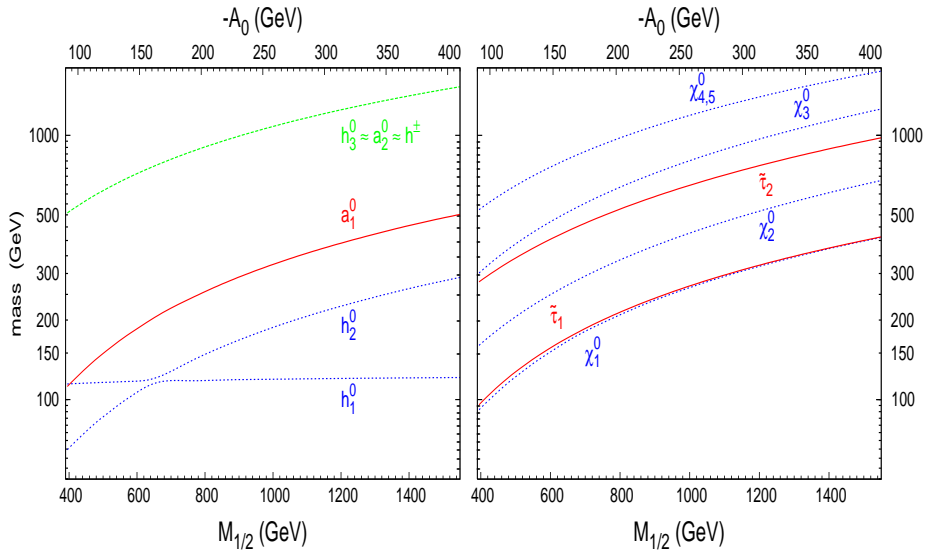

\begin{center}
\mbox{
\includegraphics[width=7.2cm,height=6.cm,angle=-90]{Plot.phenoH.epsi} 
\includegraphics[width=7.2cm,height=6.cm,angle=-90]{Plot.phenoX0.epsi} 
}
\end{center}
\vspace*{-3mm}
\caption{The Higgs (left) and neutralino plus stau (right) mass spectra in GeV 
as a function of $M_{1/2}$ along the DM line; the values of $A_0$ are  
indicated in the upper axis.}
\end{figure}

The right-hand side of Fig.~2 displays the neutralino and stau spectrum, the
lighter stau $\widetilde{\tau}_1$ being the next-to-LSP with a mass $\sim 3$ to
5~GeV above the one of the $\chi_1^0$ singlino-like LSP. $\chi_2^0$ and
$\chi_3^0$ are, respectively,  bino and wino-like while  the nearly degenerate
$\chi_{4,5}^0$ states are Higgsino-like. The charginos $\chi_1^\pm$ and
$\chi_2^\pm$ are nearly degenerate in mass with, respectively, $\chi_3^0$ and
$\chi_{4,5}^0$ (notice that the mass of the latter is $\sim \mu_{\rm eff}$). 
The remaining sparticle spectrum is very ``cMSSM''-like and can be obtained by
running the program NMSPEC \cite{nmssmtools} with input parameters as in Fig.~3
below (and $m_0 \approx 0$) and also by using any cMSSM-based  code, since the
singlet sector practically decouples from the SUSY spectrum. One  approximately
obtains $m_{\tilde g} \approx m_{\tilde q} \approx 2 M_{1/2}$  for the  gluino
and  (first/second generation) squark masses.\s

Let us now discuss the impact of other values of the parameters $m_0$ and
$\lambda$. As already stated above, the Higgs and sparticle  spectra change
very  little with $\lambda$ provided that $\lambda$ remains small enough.  Upper
bounds on $\lambda$ result from LEP constraints on Higgs scalars with masses
below the SM limit of $\sim 114$~GeV. For $M_{1/2} \lsim 660$~GeV, increasing
$\lambda$ increases the mixing of the singlet-like CP-even scalar with
doublet-like CP-even scalars and hence its couplings to the $Z$ boson, which
must not be too large. For $M_{1/2} \gsim 660$~GeV, a stronger mixing among the
CP-even scalars can lower the mass of the lighter Higgs boson which is now
SM-like, until it violates the LEP bound.\s

\begin{figure}[!ht]
\begin{center}
\includegraphics[width=.5\linewidth,angle=-90]{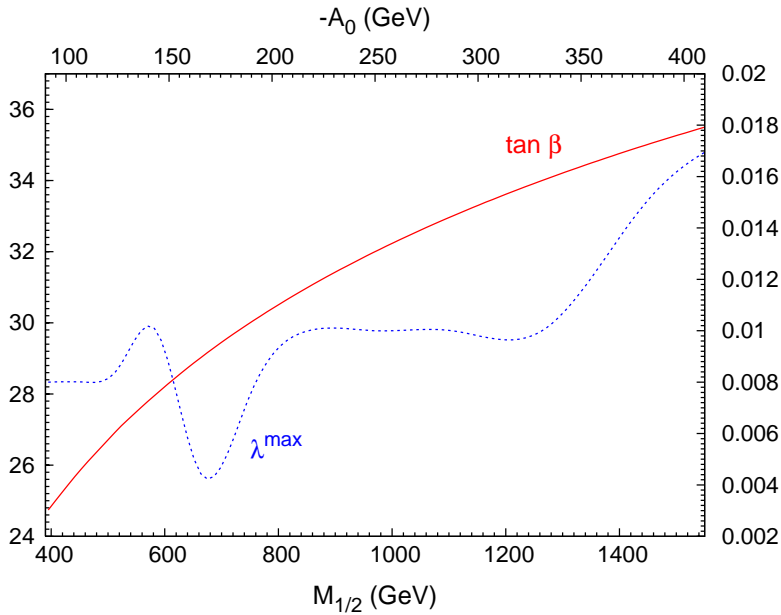} 
\end{center}
\vspace*{-5mm}
\caption[]{$\tan\beta$ and the upper bound on $\lambda$ as a function
of $M_{1/2}$ along the DM line.}
\end{figure}

Fig.~3 shows the corresponding upper limits on the parameter $\lambda$, which
are particularly strong in the ``cross\-over''-region near $M_{1/2} \sim
660$~GeV,  where relatively small values of $\lambda$ can generate a large
mixing angle; in  all cases, one has $\lambda \lsim 0.02$.  For completeness, we
also show the values of $\tan\beta$ along the DM  line.\s

We have also investigated the cNMSSM parameter space for non-zero $m_0$ and
found regions which comply with all  phenomenological constraints, including
the  DM relic density. However,  in order to generate a non-vanishing vev of the
singlet, the inequality $ m_0\lsim \frac13 |A_0|$ has to hold \cite{genNMSSM2}
and, thus, solutions  only exist for small $m_0$. Even for  $M_{1/2} \approx
1.5$ TeV, i.e. for  squark and gluino with masses of order 3 TeV and already
beyond the LHC reach, only values for $m_0$  below $\approx 140$ GeV are
allowed. Smaller $M_{1/2}$ corresponds to smaller $A_0$ and, for instance, only
$m_0 \lsim 20$~GeV is viable for $M_{1/2} \approx 400$~GeV.  Since $m_0 \ll
M_{1/2}$ in all cases, we expect the phenomenology of the model to be similar
to  the one discussed here. More details will be given in a forthcoming 
publication~\cite{preparation}.\s

Finally, let us note that we have also verified possible instabilities of the
potential along charge and colour breaking directions in field space, which
could be dangerous for $m_0 \approx 0$\cite{genNMSSM2,ccbnmssm}.  Using
analytic  approximations to the potential along such dangerous directions, the
inequality $\left(A_0-\frac12 M_{1/2}\right)^2 \lsim 9\, m_0^2+\frac83 \,
M_{1/2}^2$ was derived in Ref.~\cite{genNMSSM2}, and is satisfied within the
allowed region. Concerning  directions in field space in which the  potential is
unbounded from below, it was shown in Ref.~\cite{abelsavoy} that even if they
exist, the decay rate of the standard vacuum is usually much larger than the age
of the Universe.\s

In conclusion, we have shown that the NMSSM with universal boundary conditions
at the GUT scale is a very constrained scenario. For small values of the
universal scalar mass $m_0$, which are theoretically interesting and excluded
in  the cMSSM, all present collider constraints on sparticle and Higgs masses
are satisfied. Moreover, the requirement of a correct relic density for the 
dark matter candidate further constrains the parameter space to a
one-dimensional $[M_{1/2},A_0]$ line. Thus, only one single parameter, which can
be taken as  $M_{1/2}$,  is required to describe the  salient features of the
model.\s 

This  model leads to an interesting phenomenology. For large $M_{1/2}$, the
lightest  CP-even Higgs boson is SM-like with a  mass smaller than $\sim
120$~GeV. For small $M_{1/2}$, the lightest CP-even Higgs boson is a possibly
very light singlet-like state which will, however, be very difficult to detect
given the small value  of $\lambda$. In the SUSY  sector, the singlino-like LSP
will considerably modify the decay chains of sparticles \cite{singdecay}: one
expects that {\it all} sparticles decay via the lightest tau slepton which then
decays into the singlino-like LSP,  leading to missing energy. For very small
$\lambda$, the lifetime of the tau slepton can become so large that its track
can be visible \cite{singdecay}. In any case sparticle decays will differ in a
spectacular way from the ones expected within MSSM-typical scenarios, which will
allow to test the present scenario hopefully in the near future at the LHC.
\bigskip

{\bf Acknowledgments:}  AD is grateful to the Leverhulme Trust (UK) and to  the
Alexander von-Humboldt Foundation (Germany). We acknowledge  support from the
French ANR project PHYS@COL\&COS and discussions with S.F. King and S.
Moretti.


\begin{thebibliography}{999} 

\bibitem{genNMSSM1} H.P. Nilles, M. Srednicki and D. Wyler, Phys. Lett. B
\textbf{120} (1983) 346; J.M. Frere, D.R. Jones and S. Raby, Nucl. Phys. B
\textbf{222} (1983) 11; J. Ellis, J. Gunion, H. Haber, L. Roszkowski and F.
Zwirner, Phys. Rev. D \textbf{39}  (1989) 844; M. Drees, Int. J. Mod. Phys. A
\textbf{4}  (1989) 3635.

\bibitem{genNMSSM2} U. Ellwanger, M. Rausch de Traubenberg and C.A. Savoy, Phys.
Lett. B \textbf{315} (1993) 331; Z. Phys. C {\bf 67} (1995) 665 and Nucl. Phys.
B \textbf{492} (1997) 307.

\bibitem{genNMSSM3}   T.~Elliott, S.F.~King and P.~White, Phys.\ Lett.\  B {\bf
351} (1995) 213; S.F. King and P. White, Phys. Rev. D \textbf{52} (1995) 4183.

\bibitem{MuProblem}  J.E.~Kim and H.P.~Nilles, Phys.\ Lett.\ B {\bf 138} (1984)
 150.

\bibitem{benchmark} For a recent discussion and more references, see A. Djouadi
et al.,  arXiv:0801.4321 [hep-ph].

\bibitem{msugra} A.H. Chamseddine, R. Arnowitt and P. Nath, Phys. Rev. Lett.
{\bf 49} (1982) 970; R. Barbieri, S. Ferrara and C. Savoy, Phys. Lett. B {\bf 
119} (1982) 343; L. Hall, J. Lykken and S. Weinberg, Phys. Rev. D {\bf 27}
(1983) 2359. 

\bibitem{noscale}A.~B.~Lahanas and D.~V.~Nanopoulos, Phys.\ Rept.\  {\bf 145}, 1
  (1987); N.~Dragon, U.~Ellwanger and M.~G.~Schmidt, Prog.\ Part.\ Nucl.\
  Phys.\  {\bf 18} (1987) 1.

\bibitem{pdg} Particle Data Group [W.-M. Yao et al.], J. Phys. G {\bf 33} (2006)
1.


\bibitem{nmssmtools}  U.~Ellwanger, J.~F.~Gunion and C.~Hugonie, JHEP {\bf 0502}
(2005) 066;  U.~Ellwanger and C.~Hugonie, Comput.\ Phys.\ Commun.\  {\bf 175}
(2006) 290 and Comput.\ Phys.\ Commun.\  {\bf 177} (2007) 399; (see also the web
site {\sf http://www.th.u-psud.fr/NMHDECAY/nmssmtools.html}).

\bibitem{belsemenov} G.~Belanger, F.~Boudjema, C.~Hugonie, A.~Pukhov and
A.~Semenov, JCAP {\bf 0509} (2005) 001; G.~Belanger, F.~Boudjema, A.~Pukhov and
A.~Semenov, Comput.\ Phys.\ Commun.\  {\bf 174} (2006) 577.

\bibitem{hugbelpuk}  C.~Hugonie, G.~Belanger and A.~Pukhov, JCAP {\bf 0711}
  (2007) 009.

\bibitem{bphys} G.~Hiller, Phys.\ Rev.\  D {\bf 70} (2004) 034018;  F.~Domingo
and U.~Ellwanger, JHEP {\bf 0712} (2007) 090.

\bibitem{g-2} G. Bennett {\it et al.}, Phys. Rev. D {\bf 73} (2006) 072003; for 
a recent review see Z. Zhang, arXiv:0801.4905 [hep-ph].

\bibitem{wmap}  D.~N.~Spergel {\it et al.}  [WMAP Collaboration], Astrophys.\
J.\ Suppl.\  {\bf 170} (2007) 377.


\bibitem{largetb}  B.~Ananthanarayan and P.N.~Pandita, Phys.\ Lett.\  B {\bf
 353} (1995) 70; Phys.\ Lett.\  B {\bf 371} (1996) 245;  Int.\ J.\ Mod.\ Phys.\
 A {\bf 12} (1997) 2321.

\bibitem{ccbnmssm} U. Ellwanger and C. Hugonie, Phys. Lett. B {\bf 457} (1999)
 299.

\bibitem{abelsavoy}  S.~A.~Abel and C.~A.~Savoy, Nucl.\ Phys.\  B {\bf 532}
  (1998) 3.

\bibitem{preparation} A. Djouadi, U. Ellwanger and A.M. Teixeira, in progress. 

\bibitem{singdecay} F.~Franke and H.~Fraas, Z.\ Phys.\  C {\bf 72} (1996) 309;
U.~Ellwanger and C.~Hugonie, Eur.\ Phys.\ J.\  C {\bf 5} (1998) 723 and  Eur.\
Phys.\ J.\  C {\bf 13} (2000) 681; V.~Barger, P.~Langacker and G.~Shaughnessy,
Phys.\ Lett.\  B {\bf 644} (2007) 361 and Phys.\ Rev.\  D {\bf 75} (2007)
055013. 

\end{thebibliography}
\end{document}